# Effects of temperature gradient on the interface microstructure and diffusion of diffusion couples: phase-field simulation[*]

Li Yong-Sheng (李永胜)[†] Wu Xing-Chao (吴兴超), Liu Wei (刘苇),

Hou Zhi-Yuan (侯志远), and Mei Hao-Jie (梅浩杰)

School of Materials Science and Engineering, Nanjing University of Science and Technology, Nanjing 210094, China

**Abstract**

The temporal interface microstructure and diffusion in the diffusion couples with the mutual interactions of temperature gradient, concentration difference and initial aging time of the alloys were studied by phase-field simulation, the diffusion couples are produced by the initial aged spinodal alloys with different compositions. Temporal composition evolution and volume fraction of the separated phase indicates the element diffusion direction through the interface under the temperature gradient. The increased temperature gradient induces a wide single-phase region at two sides of the interface. The uphill diffusion proceeds through the interface, no matter the diffusion directions are up or down to the temperature gradient. For an alloy with short initial aging time, phase transformation accompanying the interdiffusion results in the straight interface with the single-phase regions at both sides. Comparing with the temperature gradient, composition difference of diffusion couple and initial aging time of the alloy show greater effect on the diffusion and interface microstructure.

**PACS**：64.60.-i, 64.75.Nx, 66.30.Fq, 68.35.bd

---

[*] Project supported by the National Natural Science Foundation of China (Grant No.51001062) and the Fundamental Research Funds for the Central Universities (Grant No.309920130121012).
[†] Corresponding author. E-mail: ysli@njust.edu.cn

# 1. Introduction

The interdiffusion between the layers of the metallic multilayer can change the interface microstructure as temperature increases, which happens in the component of alloy coatings, the interconnection of the packaging structures and multilayer thin films.[1-4] From surface to inner, the temperature gradient arises from the temperature difference of layers, the element interdiffusion through the interface and the interface microstructure will be changed thereby. As a result, the mechanical and physical properties of the multilayer metal are affected. Numerous studies have been performed on the interface reaction and interdiffusion layers metals,[5-11] in which the diffusion couple is widely used as the model to study the interdiffusion and microstructure of multilayer metal.

The diffusion path and microstructure evolution in the isothermal and stress-free multilayer diffusion couples formed by the model ternary alloy with miscibility gap have been studied by Wang et al.[12] using the phase-field simulation. Sohn and Dayananda[13] studied the double-serpentine diffusion path in annealed Fe-Ni-Al alloys, the average values of ternary interdiffusion coefficients were determined and employed to model the concentration profiles. As an important factor, the temperature gradient[14-18] has been studied in the bulk solid system for the diffusion and microstructure evolution. Hofman et al.[14] calculated the constituent redistribution driven by the temperature gradient based on thermal diffusion mechanism model. Hu et al.[15] investigated the void diffusion under an applied temperature gradient field by using the Cahn-Hilliard diffusion equation, they demonstrated that the void migrate up the temperature gradient. Sohn et al.[16] devised a diffuse interface model to simulate the effect of temperature gradient on the composition and microstructure of stress free binary alloy, their heat transport model indicates that both atomic mobility and heat transport affect the magnitude and direction of flux. Snyder et al.[17] simulated the Ostwald ripening under the temperature gradient by using the steady-state diffusion equation, they found that the volume fraction of particles decreases at high temperature regions and increases at low temperature regions, the coarsening kinetics also be affected by the temperature gradient. Ta et al.[18] studied the effect of temperature gradient on the microstructure evolution in various Ni-Al-Cr bond coat/substrate systems with thermodynamic and atomic mobility, the phase-field simulation showed that the temperature gradient promotes diffusion of both Al and Cr, which greatly accelerates the failure of various bond coat/substrate systems.

As concluded in the previous studies, the temperature gradient has a great effect on the diffusion and microstructure, it is theoretically important to investigate the effects of temperature gradient on the interdiffusion and interface microstructure of multilayer metals. However, due to the experimental complex on study the temperature gradient in multilayer metals with thickness from nano- to micro-scale, the effects of temperature gradient on the diffusion and microstructure of layer metals are less investigated.

Therefore, this paper is focused on the effects of temperature gradient on the interface microstructure and diffusion through the interface of the diffusion couple, the mutual interactions of composition difference and initial aging time (IAT) of alloy in the diffusion couple are discussed with the temperature gradient. The phase-field simulation[11,12,15,16,18-22] as a favorable method is adopted to study the interdiffusion and interface microstructure of diffusion couples under temperature gradient. The linear temperature gradient was applied in the diffusion couple formed by the aged alloy with different initial aging times and different compositions including B-enriched $\alpha_1$ and A-enriched $\alpha$ phases in the spinodal alloy, the element diffusion direction and interface microstructure will be investigated under the temperature gradient by the phase-field simulation.

## 2. Model and methods

The microstructure evolution in the diffusion couples can be described by the composition field $c$ as a function of position $r$ and time $t$, the temporal composition is controlled by the Cahn-Hilliard equation[23]

$$\frac{\partial c(\mathbf{r},t)}{\partial t} = \nabla \cdot \left[ M(c) \nabla \left( \frac{\delta G}{\delta c} - \kappa \nabla^2 c \right) \right]. \tag{1}$$

In where, $M(c)$ is chemical mobility, $G$ is the molar Gibbs energy, $\kappa$ is the gradient energy coefficient.

The chemical mobility $M(c)$ is given by the Darken equation[24]

$$M(c) = [cM_A + (1-c)M_B]c(1-c), \qquad (2)$$

where $M_A$ and $M_B$ are the atomic mobility of components A and B, respectively. They are related with the diffusivity through the Einstein relation $M_i = D_i/RT$, where $i$ denotes the components A or B of the alloy, $R$ is the gas constant and $T$ is the absolute temperature, $D_i$ is the diffusion coefficient and is assumed a small $D_A$ and large $D_B$, they are chosen as $D_A = 5.2 \times 10^{-10} \exp(-145\text{kJmol}^{-1}/RT)$ m$^2$s$^{-1}$ and $D_B = 2.6 \times 10^{-8} \exp(-130\text{kJmol}^{-1}/RT)$ m$^2$s$^{-1}$, respectively.

The molar Gibbs energy is expressed by

$$G = RT[c \ln c + (1-c)\ln(1-c)] + L_{A,B}c(1-c), \qquad (3)$$

where $L_{A,B}$ is a regular solution parameter describing the interactions between components A and B and chosen as 9.5 kJmol$^{-1}$.

The linear temperature gradient is introduced into the system with temperature increment $\Delta T$ along the horizontal direction, the temperature $T$ at position $x_i$ is expressed by $T = T_0 + x_i\Delta T$, where $T_0$ is the initial temperature at the center of simulation cell, $\Delta T$ is the temperature increment per grid length $\Delta x = 0.1\,\mu\text{m}$, the length scale is chosen to ensure a reasonable temperature gradient distribution and the diffusion interface.[25] The temperature gradient was activated through the temperature related free energy and the diffusion coefficients in the phase-field simulation.

In order to solve the Eq. (1) numerically, the dimensionless was proceed by using the parameters $\boldsymbol{r}^* = \boldsymbol{r}/l$, $t^* = tD/l^2$, $M^* = RT_c M(c)/D$, $\kappa^* = \kappa/RT_cl^2$, $G^* = G/RT_c$, where $D = 10^{-23}$ m$^2$s$^{-1}$ is a normalization factor of the diffusion coefficient and $l$ is the length scale, $T_c = 650$ K is the critical temperature of phase decomposition of the alloy. Then the kinetic evolution equation Eq. (1) is written as

$$\frac{\partial c(\boldsymbol{r}^*, t^*)}{\partial t^*} = \nabla^* \cdot \left[ M^* \nabla^* \left( \frac{\delta G^*}{\delta c} - \kappa^* (\nabla^*)^2 c \right) \right]. \qquad (4)$$

The dimensionless simulation cell is chosen as $512\,\Delta x^* \times 64\,\Delta y^*$. A thermal fluctuation [-0.005, 0.005] was added to the initial composition to trigger the phase transformation, the gradient coefficient is chosen as $\kappa^* = 0.5$. The Eq. (4) was solved by using the semi-implicit Fourier spectrum method[26] with time step $\Delta t^* = 0.02$. The lattice mismatch is chosen to be a very small value in the alloy, so the elastic strain is ignored in the present simulation.

## 3. Results and discussion

The calculation was performed with the periodic boundary condition in a symmetric diffusion couple, where the temperature increases from centre $T_0$ to boundary by $\Delta T$ along the horizontal direction in the simulation cell, as signed by the arrows in Fig. 1. Therefore, we can study one of the diffusion couple with $x^*$=1-256. The diffusion couple was formed by the two alloys aged for time $t_0^*$ with initial average composition $c_0^1$ and $c_0^2$ of the component B at each side of the diffusion couples, as shown in Fig. 1. Then the diffusion couples were annealed for time $t^*$. In the simulated figures, the red and blue regions delegate the B-enriched $\alpha_1$ phase and A-enriched $\alpha$ phase, respectively, as signed by the arrows in Fig.1(c). The color bars indicate the composition value.

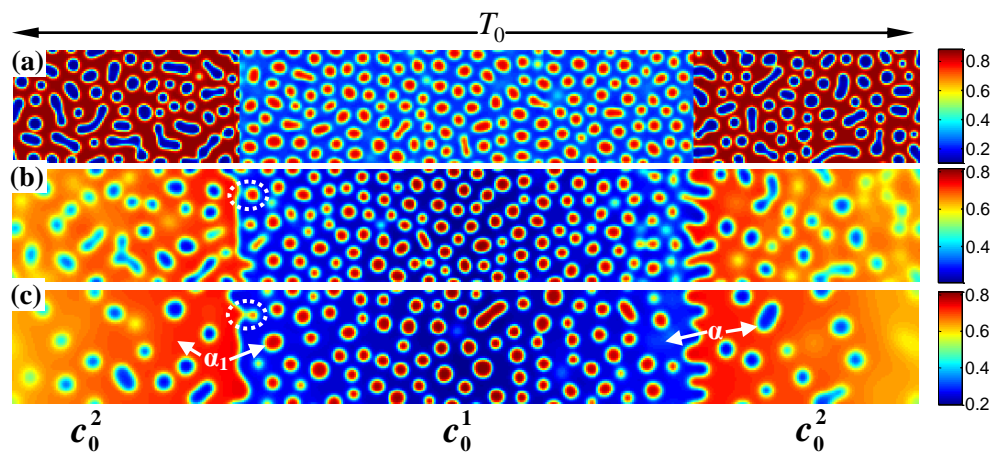

（Color online）**Fig. 1.** The microstructure evolution of diffusion couples with initial compositions $c_0^1 = 0.42$ and $c_0^2 = 0.58$ under the temperature gradient $\Delta T$=0.3 K/μm for $T_0$=470 K and $t_0^* = 160$, (a) $t^*$=0, (b) $t^*$=160, (c) $t^*$=480.

*3.1 Interface microstructure and diffusion under temperature gradient*

The microstructure evolution of diffusion couples with initial composition $c_0^1 = 0.42$ and $c_0^2 = 0.58$ (atom fraction of element B) is presented in Fig. 1, where the diffusion couple was annealed for $t^* = 480$ under the temperature gradient $\Delta T$=0.3 K/μm for $T_0$=470 K, the initial aging time of the alloys is $t_0^* = 160$.

Fig. 1(a) shows the initial microstructure of the diffusion couple at $t^*$=0 with the distinct interface. As the annealing progresses, the separated α phases are dissolved at the high temperature regions, as shown in Fig. 1(c). At the same time, the $α_1$ phase merges into the $α_1$ matrix and α phase merges into α matrix of the opposite side, respectively, as signed by the dotted circles in Figs. 1(b) and 1(c). The single-phase regions at two sides of the interface become wider as the dissolving of separated particles near the interface, as shown in Figs. 1(b)-(c). In the spinodal alloy, the solute undergoes the uphill diffusion driven by the chemical potential, the solute B at the side $c_0^1 = 0.42$ diffuse to the side $c_0^2 = 0.58$ with $α_1$ matrix, results in the merged matrix of $α_1$ and α at the interface region.

When the temperature gradient is introduced, the element diffusion through the interface will change the microstructure more obviously. The interface microstructure of diffusion couples annealing for $t^*$=600 with temperature gradients $\Delta T$=0.0, 0.2 and 0.3 K/μm are showed in Figs. 2(a)-(c), respectively. The initial composition and aging time of the diffusion couples in Fig. 2 are same to those of Fig. 1. As the temperature gradient increases, the fast diffusion results in the dissolution of $α_1$ phase near the interface at the side $c_0^1 = 0.42$, such as the $α_1$ phase particles signed with the arrows in Fig. 2. The α phase particles near the interface are shrunk or dissolved at the side $c_0^1 = 0.58$. As a result, the single-phase regions beside the interface become wider. Simultaneously, the α phases are dissolved at the high temperature regions for the large temperature gradient $\Delta T$=0.3 K/μm, as shown in Fig. 2(c).

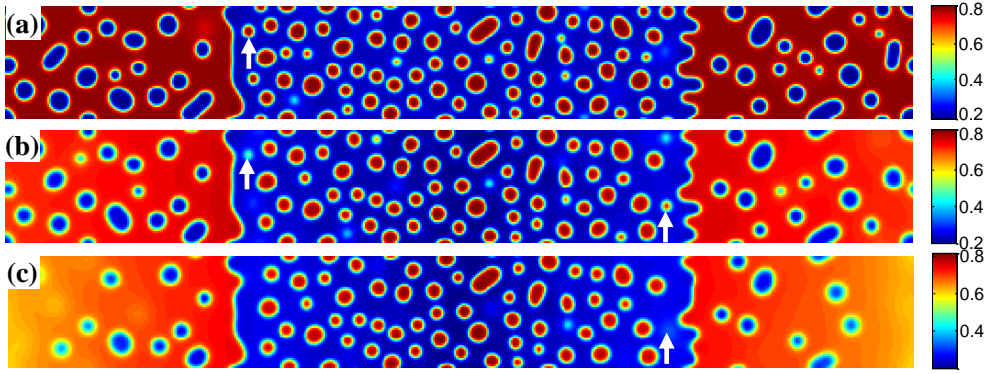

（Color online）**Fig. 2.** The microstructure of diffusion couples for alloys with initial compositions $c_0^1 = 0.42$ and $c_0^2 = 0.58$ annealing for $t^*$=600 under the temperature gradient $\Delta T$=0.0, 0.2 and 0.3 K/μm in (a)-(c), respectively. $T_0$=470 K, $t_0^* = 160$.

As a phase transformation driven by thermodynamic, the B-enriched $\alpha_1$ and A-enriched $\alpha$ are coexistence in the diffusion couples, the element diffusion through the interface and the dissolution of the initial separated phases are both related with the thermodynamic phase equilibrium. Due to the temperature change continuously in the diffusion couple, the equilibrium composition of $\alpha_1$ phase decreases and $\alpha$ increases as the temperature increases, it is just like the Soret effect[16] that induces the composition gradient in the diffusion couples, then the kinetic diffusion is affected by the composition gradient. Therefore, the concentration of elements A and B at two sides of the diffusion couples can be changed by the diffusion, which will be discussed following by the quantitative concentration variation.

The temporal average compositions $\langle c_B^{t^*} \rangle$ of element B at two sides $x^*$=1-128 ($c_0^2 = 0.58$) and $x^*$=129-256 ($c_0^1 = 0.42$) of the diffusion couples are plotted in Fig. 3. For different temperature gradients, the average compositions increase at the side with high concentration $c_0^2 = 0.58$, and they decrease at the side with low concentration $c_0^1 = 0.42$. It can be concluded that the element B diffuses from the low concentration side to high concentration side, i.e. the uphill diffusion through the interface. Also, we can know from the composition conservation that the diffusion direction of element A is inverse to that of element B.

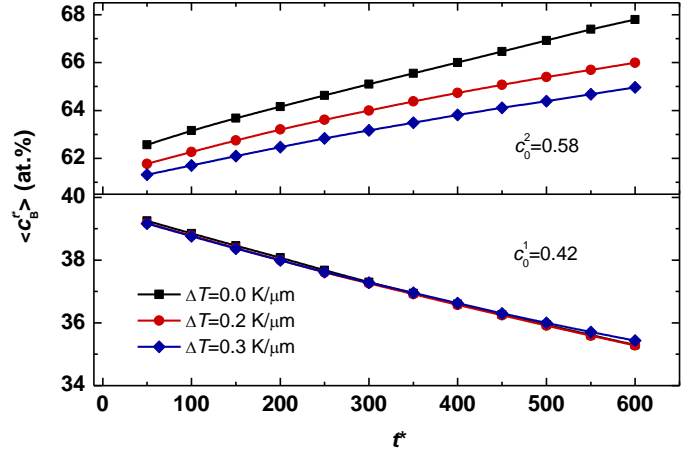

(**Color online**) **Fig. 3.** Temporal average composition $\langle c_B^{t^*} \rangle$ at two sides $x^*$=1-128 ($c_0^2 = 0.58$) and $x^*$=129-256 ($c_0^1 = 0.42$) of diffusion couples showed in Fig. 2.

In addition, the values of $\langle c_B^{t^*} \rangle$ are small for the large $\Delta T$ at the side $c_0^2 = 0.58$, while they are almost the same at the side $c_0^1 = 0.42$ for $\Delta T$=0.0, 0.2 and 0.3 K/μm. As the thermal transport is ignored in the model, the decrease of $\langle c_B^{t^*} \rangle$ for a large $\Delta T$ at the high temperature is caused by the changed equilibrium composition of the alloy, i.e. the higher the temperature, the lower the equilibrium composition of α$_1$ phase. The difference of the $\langle c_B^{t^*} \rangle$ before and after annealing is calculated by $|\Delta c| = \left| \langle c_B^{t_2^*} \rangle - \langle c_B^{t_1^*} \rangle \right|$, where $\langle c_B^{t_1^*} \rangle$ and $\langle c_B^{t_2^*} \rangle$ are the average composition of component B at $t^*$=0 and $t^*$=600, respectively. The average composition difference $|\Delta c|$ for different temperature gradients and IATs are listed in Table 1. It is showed that the $|\Delta c|$ is larger for $t_0^* = 40$ than that of $t_0^* = 160$ for different temperature gradients, therefore, the interdiffusion of elements through the interface of diffusion couple is intensified for a short IAT. The variation of $|\Delta c|$ also indicates the great $\Delta T$ weakens the diffusion of element B to the high temperature regions.

**Table 1** Average concentration difference $|\Delta c|$ (at.%) at the left ($c_0^2 = 0.58$) and right sides ($c_0^1 = 0.42$) of the diffusion couples annealing for $t^*$=600 showed in Fig. 3 and Fig. 6.

|  | $t_0^* = 160$ | | | $t_0^* = 40$ | | |
|---|---|---|---|---|---|---|
| $\Delta T$ (K/μm) | 0.0 | 0.2 | 0.3 | 0.0 | 0.2 | 0.23 |
| $c_0^2 = 0.58$ | 5.2 | 4.2 | 3.6 | 9.8 | 9.8 | 8.5 |
| $c_0^1 = 0.42$ | 3.9 | 3.8 | 3.7 | 5.2 | 5.3 | 5.3 |

The variations of B-enriched $α_1$ phase volume fractions at the region $x^*$=129-256 ($c_0^1 = 0.42$) and $y^*$=1-64 in Fig. 2 are plotted as a function of time, the volume fractions for $t_0^* = 40$, $\Delta T$=0.0 and 0.23 K/μm are also given together, as shown in Fig. 4. The volume fractions decrease as annealing progresses, the larger the temperature gradient, the smaller the volume fraction for the same annealing time. The volume fractions decrease at high temperature regions was also demonstrated in the temperature gradient system studied by Snyder et al.[17]

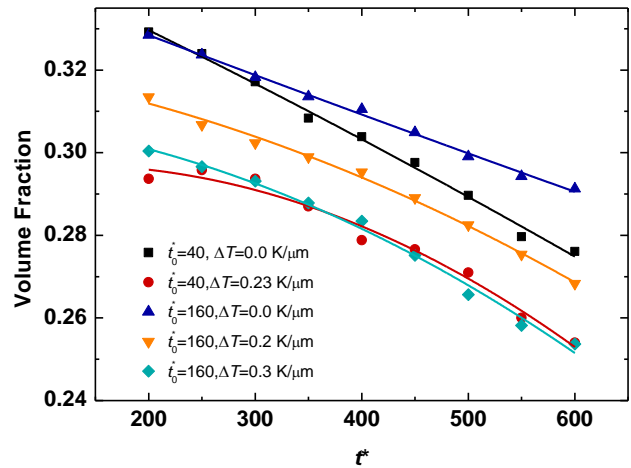

(**Color online**) **Fig. 4.** Volume fraction variations of B-enriched $α_1$ phase in the diffusion couples with $x^*$=129-256 and $y^*$=1-64 for different initial aging times and temperature gradients showed in Fig. 2 and Fig. 5.

The reduced volume fraction is caused by two reasons in the temperature gradient diffusion couples. First, the increased temperature reduces the supercooling, i.e. the driving force of the phase separation is decreased at high temperature, as a result, the phase separation is retarded or the separated phases are dissolved again by annealing. Second, the decrease of volume fractions is the departure of element B from the low concentration side $c_0^2 = 0.42$ to the high concentration side $c_0^2 = 0.58$ during the annealing, the variation of volume fraction is consistent with the composition decrease (Fig. 3) and the recession of α1 phase near the interface (Fig. 2).

*3.2 Interface diffusion and microstructure for a short initial aging time*

In Fig. 4, the volume fractions for a short IAT $t_0^* = 40$ and $\Delta T$=0.23 K/μm have the similar values as that of $t_0^* = 160$ and $\Delta T$=0.3 K/μm, as labelled by the circle and diamond in Fig. 4. The variations of volume fractions also demonstrate that the short IAT can induce the strong interdiffusion in the diffusion couples. For the short IAT, the phase separation still progresses during annealing, there are more atoms randomly distributed rather than clustering, which favors the long range diffusion of atoms across the interface of diffusion couple. As discussed in Section 3.1, the temperature gradient can induce the composition difference, a lower concentration of B-enriched α₁ and higher concentration of A-enriched α at higher temperature, the concentration difference in the diffusion couples promotes the element diffusion.

The morphology evolutions for IAT $t_0^* = 40$ are showed in Fig. 5 with different Δ*Ts*. In where, the increased temperature reduces the B element concentration at the side $c_0^2 = 0.58$, inversely, the uphill diffusion supports the B element diffuse to the side $c_0^2 = 0.58$. Therefore, the diffusion direction of B element is decided by the mutual interactions of temperature gradient and initial composition difference.

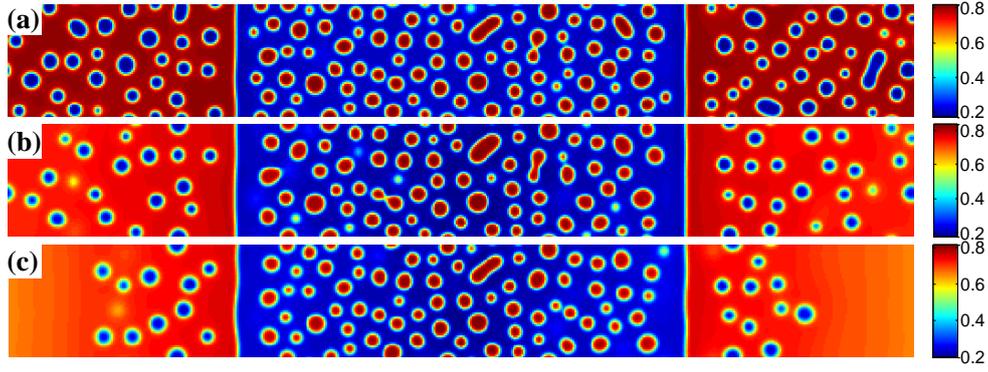

(**Color online**) **Fig. 5.** The microstructure of diffusion couples for alloys with initial compositions $c_0^1 = 0.42$ and $c_0^2 = 0.58$ annealing for $t^*=600$ under the temperature gradient $\Delta T$=0.0, 0.2 and 0.23 K/μm in (a)-(c), respectively. $T_0$=470 K, $t_0^* = 40$.

Besides the change of volume fraction with the IAT, the straight interface of the diffusion couples is clearly presented for the short IAT $t_0^* = 40$, as shown in Fig. 5. For a short IAT, there is no stable separated phase or only small size particles. As the element diffuses through the interface, the solute concentration is decreased near the interface, there is no enough driving force for the phase separation or particle growth during the annealing, so the same phase merging can not happen, which leads to the single-phase region and the straight interface in the diffusion couple.

Fig. 6 plots the temporal average composition $\langle c_B^{t^*} \rangle$ at two sides $x^*$=1-128 ($c_0^2 = 0.58$) and $x^*$=129-256 ($c_0^1 = 0.42$) of the diffusion couples showed in Fig. 5. The variations of $\langle c_B^{t^*} \rangle$ at both sides of the diffusion couple are similar to those of Fig. 3. While the decrease of $|\Delta c|$ is not obvious than that of Fig.3 as the $\Delta T$ increases at the side $c_0^2 = 0.58$.

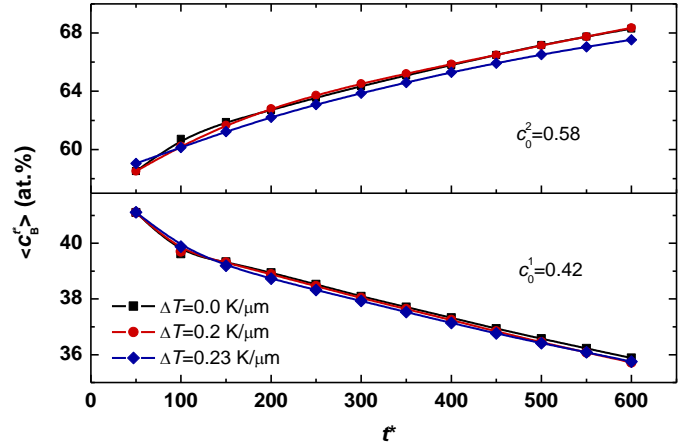

(**Color online**) **Fig. 6.** Temporal average composition $\langle c_B^{t^*} \rangle$ at two sides $x^*$=1-128 ($c_0^2 = 0.58$) and $x^*$=129-256 ($c_0^1 = 0.42$) of diffusion couples showed in Fig. 5.

Figs. 7(a) and 7(b) show the diffusion couple with $c_0^1 = 0.58$ and $c_0^2 = 0.42$ under the temperature gradient $\Delta T$=0.0 and 0.4 K/μm for $T_0$=470 K, respectively, the initial aging time is $t_0^* = 40$ and the annealing time is $t^*$=600. Here the temperature also increases from the center to boundary along the horizontal direction, while the initial compositions at the two sides of the diffusion couple are inversed to that of Fig. 5. It can be seen that the interface is also straight for the short IAT, and the single-phase α and α₁ are presented at the sides with $c_0^2 = 0.42$ and $c_0^1 = 0.58$, respectively.

The temporal average compositions $\langle c_B^{t^*} \rangle$ in the diffusion couple showed in Fig.7 are plotted in Fig. 8. The $\langle c_B^{t^*} \rangle$ still increases at the side with high concentration $c_0^1 = 0.58$ and decreases at the side with low concentration $c_0^2 = 0.42$ for $\Delta T$=0.0 K/μm and $\Delta T$=0.4 K/μm, which means that the element B diffuses from the low concentration to high concentration even if the diffusion direction is inverse to the direction of temperature increment. Therefore, the diffusion direction of elements in the diffusion couples formed by the spinodal alloys is dominated by the initial composition, the uphill diffusion progresses through the interface no matter the diffusion directions are up or down to the direction of temperature gradient.

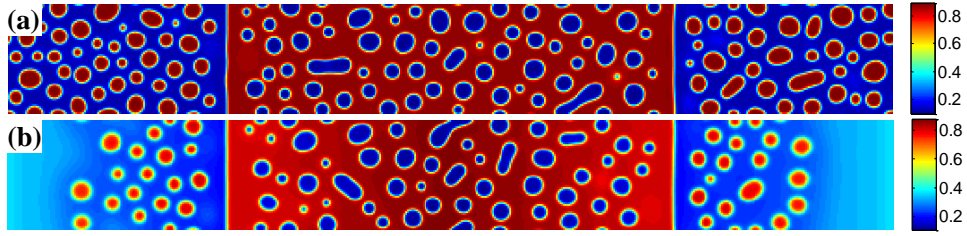

（**Color online**）**Fig. 7.** The microstructure of diffusion couples for alloys with initial compositions $c_0^1 = 0.58$ and $c_0^2 = 0.42$ annealing for $t^*$=600 under the temperature gradient $\Delta T$=0.1 K/μm (a) and $\Delta T$=0.4 K/μm (b), $T_0$=470 K, $t_0^* = 40$.

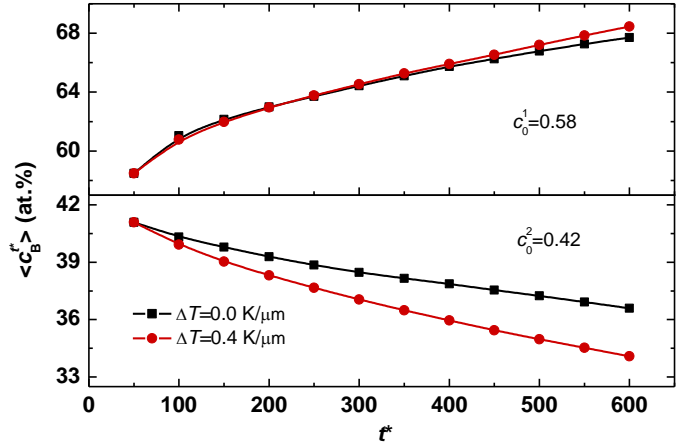

（**Color online**）**Fig. 8.** Temporal average composition $\langle c_B^{t^*} \rangle$ at two sides $x^*$=1-128 ($c_0^2 = 0.42$) and $x^*$=129-256 ($c_0^1 = 0.58$) of diffusion couples showed in Fig. 7.

*3.3 Interface diffusion and microstructure for different initial compositions*

As the initial composition difference $\Delta c_0 = c_0^2 - c_0^1$ ($c_0^2 > c_0^1$) decreases, the α phase and α$_1$ phase are connected with themselves at the interface for the long IAT $t_0^* = 160$, as shown in Figs. 9(a) and 9(c), where the initial compositions of diffusion couples are $c_0^1 = 0.45$ and $c_0^2 = 0.55$ (Fig. 9(a)), $c_0^1 = 0.48$ and $c_0^2 = 0.53$ (Fig. 9(c)), the temperature gradient is Δ$T$=0.4 K/μm. However, the single-phase regions can still be presented beside the interface for the short IAT $t_0^* = 40$, as shown in Figs. 9(b) and 9(d), which have the same initial compositions to those of Figs. 9(a) and 9(c), respectively. It can be concluded from the above results that the initial composition difference $\Delta c_0$ and IAT of the alloy in the diffusion couple have the dominant influence on the interface microstructure compared with the temperature gradient.

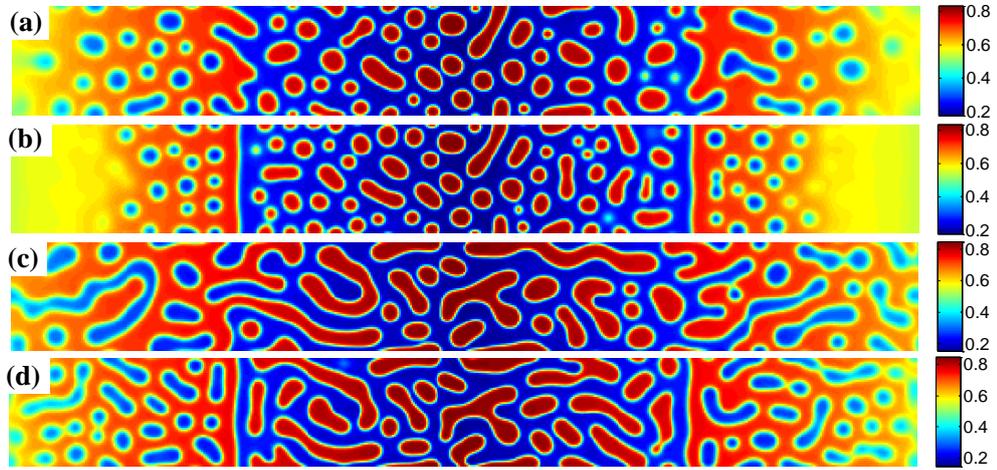

（Color online）**Fig. 9.** The microstructure of diffusion couples with initial compositions $c_0^1 = 0.45$ and $c_0^2 = 0.55$ in (a) and (b), $c_0^1 = 0.48$ and $c_0^2 = 0.53$ in (c) and (d) annealing for $t^*$=600 under the temperature gradient Δ$T$=0.4 K/μm, $T_0$=470 K. (a) and (c) $t_0^* = 160$, (b) and (d) $t_0^* = 40$.

Fig. 10 plots the temporal average composition $\langle c_B^{t^*} \rangle$ at the regions $x^*$=1-128 and $x^*$=129-256 of Figs. 9(a)-(d). It can be seen that the average compositions increase at the high concentration side and decrease at the low concentration side, the $|\Delta c|$ for $t_0^* = 40$ are larger than that of $t_0^* = 160$ (Table 2). As the initial composition difference $\Delta c_0$ decreases from 0.16 to 0.1 and 0.05, the composition differences $|\Delta c|$ are decreased at both sides of the diffusion couple after annealing, which can be seen from Table 1 and Table 2. It can be deduced from the composition variation that the interdiffusion is reduced as the $\Delta c_0$ decreases, which is due to the decreased driving force of diffusion comes from the composition gradient.

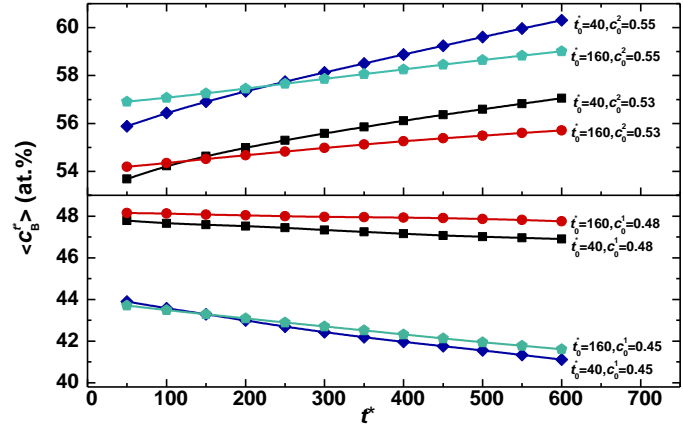

(**Color online**) **Fig. 10.** Temporal average composition $\langle c_B^{t^*} \rangle$ at the two sides $x^*=1-128$ and $x^*=129-256$ of diffusion couples showed in Fig. 9.

**Table 2** Average concentration difference $|\Delta c|$ (at.%) at the two sides of the diffusion couples annealing for $t^*=600$ showed in Fig. 10.

|  | $c_0^1 = 0.45$ | $c_0^2 = 0.55$ | $c_0^1 = 0.48$ | $c_0^2 = 0.53$ |
| --- | --- | --- | --- | --- |
| $t_0^* = 160$ | 2.1 | 2.1 | 0.4 | 1.5 |
| $t_0^* = 40$ | 2.7 | 4.4 | 0.8 | 3.3 |

The variations of volume fractions of B-enriched $\alpha_1$ phase at regions with $x^*=129-256$ and $y^*=1-64$ of Fig. 9 are plotted in Fig. 11. The volume fractions decrease from about 0.39 to 0.35 in the diffusion couple with $c_0^1 = 0.45$ and $c_0^2 = 0.55$, as showed by the lines signed with circle, while the volume fractions have almost no change in the diffusion couple with $c_0^1 = 0.48$ and $c_0^2 = 0.53$, see the lines signed with square. The variations of volume fractions demonstrate again that the smaller $\Delta c_0$ results in the weaker interdiffusion under temperature gradient.

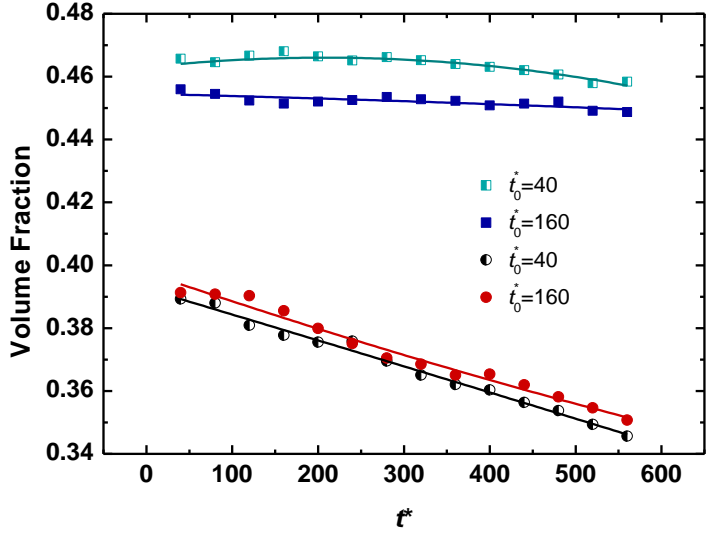

（**Color online**）**Fig. 11.** Volume fraction variations of B-enriched $\alpha_1$ phase in the diffusion couples with $x^*$=129-256 and $y^*$=1-64 of Fig. 9. The lines labelled with circle and square delegate the diffusion couple with initial compositions $c_0^1 = 0.45$ and $c_0^2 = 0.55$, $c_0^1 = 0.48$ and $c_0^2 = 0.53$, respectively.

The simulation results give an understanding on the interface microstructure and diffusion in the diffusion couple under the effects of temperature gradient, initial composition difference and different initial aging times. As the thermodynamic driving force of phase transformation, the chemical free energy changes with the temperature, thus the composition and microstructure will be changed in the temperature gradient system. In addition, the uphill diffusion dominates the diffusion direction through the interface no matter the direction of temperature gradient is. Therefore, the interface diffusion and microstructure in the spinodal alloy system show some differences to that of heat transport and mobility controlled diffusion.[16] is It is theoretical significant for the basic understanding on the interface diffusion in the diffusion couple formed by spinodal alloy.

**4. Summary and conclusions**

The effects of temperature gradient on the diffusion and interface microstructure of diffusion couples were studied in the binary spinodal alloys, the mobility as a function of temperature and temporal composition was incorporated into the Cahn-Hilliard equation, the thermodynamic and dynamic parameters were adopted for the phase transformation. By calculating the temporal composition of elements and the volume fraction of separated phase in the diffusion couples, the diffusion direction of element through the interface is clarified under temperature gradient. The evolution of interface microstructure is presented by the simulated morphology. At the same time, the effects of initial aging time and the composition difference of the alloy on the diffusion and microstructure were also studied with the temperature gradient.

The simulation results show that the interface microstructure, straight interface with single-phase or interconnected interface, are affected mainly by the initial aging time of the alloy under the temperature gradient, the element diffusion through the interface depends on the initial composition difference of two sides of the diffusion couple. The width of single-phase regions at both sides of the interface is enlarged as the temperature gradient increases, the straight interface with single-phase is presented for the short initial aging time under the temperature gradient. The solute element diffuses from the low concentration to high concentration through the interface of diffusion couple formed by spinodal alloy, no matter the diffusion directions are up or down to the temperature gradient.